\documentstyle[12pt,twoside, epsf, here, psfig]{article}
\newcounter{tony}
\newcommand{\fb}{\mbox{\boldmath{$f$}}}

\newcommand{\bgamma}{\mbox{\boldmath{$\gamma$}}}

\newcommand{\bsigma}{\mbox{\boldmath{$\sigma$}}}

\newcommand{\bOmega}{\mbox{\boldmath{$\Omega$}}}
\newcommand{\beq}{\begin{equation}}
\newcommand{\eeq}[1]{\label{eq:#1}\end{equation}}

\newcommand{\eqref}[1]{(\ref{eq:#1})}

      \newcommand{\beqn}{\begin{equation}}
      \newcommand{\eeqn}{\end{equation}}
      \newcommand{\beqna}{\begin{eqnarray}}
      \newcommand{\eeqna}{\end{eqnarray}}

\title{The Reason for the Efficiency of the Pian--Sumihara Basis}

\renewcommand{\thefootnote}{\fnsymbol{footnote}}

\renewcommand{\thefootnote}{\fnsymbol{footnote}}
\author{S. J. Childs\footnotemark[1] \\ {\small\em Department of Pure and Applied
Mathematics, Rhodes University, Grahamstown,} \\ {\small\em 6140, South
Africa} \\ \\ B. D. Reddy \\ {\small\em Department of Mathematics and
Applied Mathematics, University of Cape Town,} \\ {\small\em
Rondebosch, 7700, South Africa}}

\date{}       

\begin{document}

\maketitle
\footnotetext[1]{Corresponding author. Email: {\em schilds@iafrica.com}}
\renewcommand{\thefootnote}{\arabic{footnote}}

\begin{abstract}
\noindent {\em A logical explanation as to why the choice of 
\begin{eqnarray*}
\left[ \begin{array}{ccccc} 1 & 0 & 0 & \eta & 0 \\ 0 & 1 & 0 & 0 &
\xi \\ 0 & 0 & 1 & 0 & 0 \end{array} \right]
\end{eqnarray*}
(the Pian--Sumihara basis) as a linear basis to approximate
stress leads to greater efficiency in enhanced strain problems,
is presented. An Airy stress function and the consequent selective simplification resulting from the differentiation of an implied, single, parent approximating polynomial, are the essence of this argument.}
\end{abstract}


\section{Introduction}

Pian and Sumihara first identified the basis
\begin{eqnarray*}
\left[ \begin{array}{ccccc} 1 & 0 & 0 & \eta & 0 \\ 0 & 1 & 0 & 0 &
\xi \\ 0 & 0 & 1 & 0 & 0 \end{array} \right]
\end{eqnarray*}
as the most efficient linear basis for approximating stress in
enhanced strain problems. This observation they made more rigorous by way of a Wilson element (a perturbation of sorts).

This paper presents a logical mathematical argument for making the
same choice of basis, albeit with the wisdom of hindsight. It
attributes the greater efficiency of the basis to properties inherent
in the mathematics of the problem. The components of the stress
tensor are recognised to be related by way of an Airy stress function
and it is in this way that a fundamentally more correct
representation of the full linear basis is arrived at. By further
desiring the advantages of a two field problem, the most efficient,
linear basis is obtained. 

\section{An Airy Stress Function}

The Airy stress function is a potential of sorts. Interpreting stresses to be the various second derivatives of a single polynomial leads to selective simplification and interdependence between the resulting linear approximations. This simplification and the interdependence are not obvious in a more superficial treatment.
\begin{eqnarray*}
\mathop{\rm div}{\bsigma} = {\bf 0} \hspace{10mm} &\Rightarrow&
\hspace{10mm} \displaystyle \frac{\partial \sigma_{11}}{\partial x} +
\frac{\partial \sigma_{12}}{\partial y} = 0 \hspace{5mm} \mbox{and}
\hspace{5mm} \displaystyle \frac{\partial \sigma_{21}}{\partial x} +
\frac{\partial \sigma_{22}}{\partial y} = 0
\end{eqnarray*}
This is recogniseable as 
\[
\mathop{\rm curl} (- \sigma_{12}, \sigma_{11}, 0) = 0 \hspace{5mm} \mbox{and} \hspace{5mm} \mathop{\rm curl} (\sigma_{22}, - \sigma_{21}, 0) = 0.
\]
This, in turn, implies that $(- \sigma_{12}, \sigma_{11}, 0)$ and
$(\sigma_{22}, - \sigma_{21}, 0)$ may be interpretted as $\nabla
\alpha$ and $\nabla \beta$ respectively, without any inconsistancy in
the 
\[
\mathop{\rm curl} \nabla ( \ \cdot \ ) = 0
\]
identity.

By symmetry of $\bsigma$,
\[
\sigma_{12} = \sigma_{21} \Rightarrow \frac{\partial \alpha}{\partial
x} - \frac{\partial \beta}{\partial y}= 0
\]
and for a two dimensional problem of the type under consideration this once again implies 
\[
\mathop{\rm curl} (\beta, \alpha, 0) = 0.
\]
$(\beta, \alpha, 0)$ may therefore be interpretted as $\nabla \Phi$ without any inconsistancy in the 
\[
\mathop{\rm curl} \nabla ( \ \cdot \ ) = 0
\]
identity.

In summary, with an equation
\[
\mathop{\rm div}{\bsigma} = {\bf 0}
\]
governing the motion, in the two-dimensional case, the components of
the stress may be derived from an Airy stress function as follows
\begin{eqnarray*}
\sigma_{11} &=& \frac{\partial^2 \Phi}{\partial y^2}, \\
\sigma_{22} &=& \frac{\partial^2 \Phi}{\partial x^2}, \\
\sigma_{12} &=& \frac{\partial^2 \Phi}{\partial x \partial y},
\end{eqnarray*}
where $\Phi$ is the Airy stress function.
 
\subsection{Finite Element Approximation}

Due to approximation,
\begin{eqnarray*}
\mathop{\rm div}{\bsigma} = {\bf 0}
\end{eqnarray*}
and not the constitutive
\begin{eqnarray*}
\mathop{\rm div}{\bsigma} = {\fb}
\end{eqnarray*}
are really the equations being solved ({\sc Reddy} \cite{comreddy:1}). 

Defining a function
\begin{eqnarray*}
\phi(\xi,\eta) \equiv \Phi(x(\xi,\eta), y(\xi,\eta))
\end{eqnarray*}
on each element $\Omega_e$, 
\begin{eqnarray*}
\sigma_{22} &=& \frac{\partial^2 \Phi}{\partial x_1^2} \\ 
&=& \frac{\partial}{\partial \xi} \left( \frac{\partial
\phi}{\partial \xi} \frac{\partial \xi}{\partial x_1} +
\frac{\partial \phi}{\partial \eta} \frac{\partial \eta}{\partial
x_1}  \right) \frac{\partial \xi}{\partial x_1} +
\frac{\partial}{\partial \eta} \left(  \frac{\partial \phi}{\partial
\xi} \frac{\partial \xi}{\partial x_1} + \frac{\partial
\phi}{\partial \eta} \frac{\partial \eta}{\partial x_1}  \right)
\frac{\partial \eta}{\partial x_1} \\
&=& \left( \frac{\partial^2 \phi}{\partial \xi^2} \frac{\partial
\xi}{\partial x_1} + \frac{\partial^2 \phi}{\partial \eta \partial
\xi} \frac{\partial \eta}{\partial x_1}  \right) \frac{\partial
\xi}{\partial x_1} + \left(  \frac{\partial^2 \phi}{\partial \eta
\partial \xi} \frac{\partial \xi}{\partial x_1} + \frac{\partial^2
\phi}{\partial \eta^2} \frac{\partial \eta}{\partial x_1}  \right)
\frac{\partial \eta}{\partial x_1} 
\end{eqnarray*}

\subsubsection*{Assumption}

The individual elements, $\Omega_e$, are usually mapped to the master
element, $\hat \Omega$, with $\frac{\partial \xi}{\partial x_2}
\approx \frac{\partial \eta}{\partial x_1}\approx 0$ on average,
$\frac{\partial \xi}{\partial x_1} \approx a_1$ and $\frac{\partial
\eta}{\partial x_2}\approx a_2$, $a_1$ and $a_2$ some constants, on
average. (Alternatively it can be argued that there will be no loss
of generality or weakening of the argument if a rectangular mesh is
considered. Not allowing this simplification leads to an extremely
messy argument, a chapters long exercise in differentiation.) This
implies
\begin{eqnarray*}
\sigma_{22} &=& a_1^2 \frac{\partial^2 \phi}{\partial \xi^2}. 
\end{eqnarray*}
Similarly,
\begin{eqnarray*}
\sigma_{11} &=& a_2^2 \frac{\partial^2 \phi}{\partial \eta^2} \\
\sigma_{12} &=& \sigma_{21} \ = \ a_1 a_2 \frac{\partial^2
\phi}{\partial \xi \partial \eta}
\end{eqnarray*}

\section{The Relationship Implicit in the Linear Approximation}

Since linear approximations of $\sigma_{11}$ are to be considered,
\begin{eqnarray*}
\frac{\partial^2 \phi}{\partial \eta^2} &=& b_1 + b_2 \xi + b_3 \eta 
\end{eqnarray*}
where $b_1$, $b_2$ and $b_3$ are the relevant combining constants. This means
\begin{eqnarray} \label{1}
\phi(\xi, \eta) &=& \int_{-1}^1 \int_{-1}^1 b_1 + b_2 \xi + b_3 \eta \ d \eta d \eta \nonumber \\
&=& c_1 + c_3 \eta + \frac{1}{2} b_1 \eta^2 + \frac{1}{2} b_2 \xi \eta^2 + \frac{1}{6} b_3 \eta^3 + \eta f_1(\xi) + f_2(\xi)
\end{eqnarray}
in which the exact form of $\eta f_1(\eta) + f_2(\eta)$ remains to be determined. Similarly, approximating $\sigma_22$ as some multiple of $b_4 + b_5 \xi + b_6 \eta$ implies this very same polynomial function
\begin{eqnarray} \label{2}
\phi(\xi, \eta) &=& \int_{-1}^1 \int_{-1}^1 b_4 + b_5 \xi + b_6 \eta \ d \xi d \xi \hspace{10mm} \mbox{(\em by Airy stress function)} \nonumber \\
&=& c_1 + c_2 \xi + \frac{1}{2} b_4 \xi^2 + \frac{1}{6} b_5 \xi^3 + \frac{1}{2} b_6 \xi^2 \eta + \xi g_1(\eta) + g_2(\eta),
\end{eqnarray}
in which the exact form of $g_2(\eta)$ is determined by equation
(\ref{1}). This equation in turn specifies $f_2(\xi)$ in equation
(\ref{1}). Approximating $\sigma_{12} = \sigma_{21}$ in it's turn as as $b_7 + b_8 \xi + b_9 \eta$ implies the polynomial function
\begin{eqnarray} \label{3}
\phi(\xi, \eta) &=& \int_{-1}^1 \int_{-1}^1 b_7 + b_8 \xi + b_9 \eta \ d \xi d \eta \nonumber \\
&=& c_1 + b_7 \xi \eta + \frac{1}{2} b_8 \xi^2 \eta + \frac{1}{2} b_9 \xi \eta^2 + f_2(\xi) + g_2(\eta)
\end{eqnarray}
where $f_2(\xi)$ and $g_2(\eta)$ have already been determined by equations (\ref{2}) and (\ref{1}) respectively. This last expression for $\phi(\xi, \eta)$ also specifies the, until now undetermined, $\eta f_1(\xi)$  and $\xi g_1(\eta)$ in equations (\ref{1}) and (\ref{2}). In summary, collecting equations (\ref{1}), (\ref{2}) and (\ref{3}) together leads to the specification of an implied, single parent approximating polynomial
\begin{eqnarray*}
\phi(\xi, \eta) &=& c_1 + c_2 \xi + c_3 \eta + c_4 \xi^2 + c_5 \xi
\eta + c_6 \eta^2 + c_7 \xi^3 + c_8 \xi^2 \eta + c_9 \xi \eta^2 +
c_{10} \eta^3.
\end{eqnarray*}
Having established both the existance and nature of the relationship between the constants in what were apparently seperate linear approximations,  
\begin{eqnarray*}
\frac{\partial^2 \phi}{\partial \xi^2} &=& 2c_4 + 6c_7 \xi + 2c_8 \eta
\\
\frac{\partial^2 \phi}{\partial \eta^2} &=& 2c_6 + 2c_9 \xi + 6c_{10} \eta
\\
\frac{\partial^2 \phi}{\partial \xi \partial \eta} &=& c_5 + 2c_8 \xi
+ 2c_9 \eta
\end{eqnarray*}
can now be written where the $c_i$'s ($i = 4, \cdots 10$) are constants related to the finite element solution of the problem in question.

\subsection*{Conclusion}

The Airy stress function therefore reveals how a linear approximation
of the components of $\bsigma$ on each element really amounts to
\begin{eqnarray} \label{4}
\left[ \begin{array}{c} \sigma_{11} \\ \sigma_{22} \\ \sigma_{12}
\end{array} \right] = \left[ \begin{array}{c c c c c c c} 1 & 0 & 0 &
\eta & 0 & \xi & 0 \\ 0 & 1 & 0 & 0 & \xi & 0 & \eta \\ 0 & 0 & 1 & 0
& 0 & \eta & \xi \end{array} \right] \left[ \begin{array}{c} \cdot \\
\cdot \\ \cdot \end{array} \right] 
\end{eqnarray}
instead of the superficially more obvious
\begin{eqnarray*}
\left[ \begin{array}{c} \sigma_{11} \\ \sigma_{22} \\ \sigma_{12}
\end{array} \right] = \left[ \begin{array}{c c c c c c c c c} 1 & 0 & 0 & \xi & 0 & 0 & \eta & 0 & 0 \\ 
0 & 1 & 0 & 0 & \xi & 0 & 0 & \eta & 0 \\ 
0 & 0 & 1 & 0 & 0 & \xi & 0 & 0 & \eta \end{array} \right] \left[ \begin{array}{c} \cdot \\
\cdot \\ \cdot \end{array} \right] 
\end{eqnarray*}

\section{Eliminating the Last Two Columns}

The rank of the matrix in equation (\ref{4}) indicates that there are
still two extra columns. The equation in which $\bsigma$ is used is a
three--field problem, in which the strain, $\bgamma$, only occurs
once in a term $\bsigma \cdot \bgamma$. Choosing $\bsigma$ correctly
would reduce the problem to a two--field problem since
\begin{eqnarray*}
\int \bsigma \cdot \bgamma d \bOmega &=& 0
\end{eqnarray*}
is required  in accordance with {\sc Reddy} \cite{aruna:1}. In other words
\begin{eqnarray*} 
\bsigma \cdot \bgamma &=& \left[ \begin{array}{c c c c c c c} 1 & 0 &
0 & \eta & 0 & \xi & 0 \\ 0 & 1 & 0 & 0 & \xi & 0 & \eta \\ 0 & 0 & 1
& 0 & 0 & \eta & \xi \end{array} \right] \left[ \begin{array}{c}
\cdot \\ \cdot \\ \cdot \end{array} \right] \cdot \left[ \begin{array}{c c
c c} \xi & 0 & 0 & 0 \\ 0 & \eta & 0 & 0 \\ 0 & 0 & \xi & \eta
\end{array} \right] \left[ \begin{array}{c} \cdot \\ \cdot \\ \cdot
\end{array} \right]
\end{eqnarray*}
must always be zero. This is only certaint if the sixth and seventh columns of the stress basis are omitted.

\section{Conclusion}

An Airy stress function and consequent simplification resulting from the differentiation of an implied, single, parent, approximating polynomial are able to provide a logical explanation as to why the choice of 
\begin{eqnarray*}
\left[ \begin{array}{ccccc} 1 & 0 & 0 & \eta & 0 \\ 0 & 1 & 0 & 0 &
\xi \\ 0 & 0 & 1 & 0 & 0 \end{array} \right]
\end{eqnarray*}
(the Pian--Sumihara basis) as a linear basis to approximate
stress leads to greater efficiency in enhanced strain problems.

\nocite{aruna:1}
\nocite{fox:1}

\bibliography{piansumihara}

\begin{thebibliography}{1}

\bibitem{fox:1}
R.J. Atkin and N.~Fox.
\newblock {\em An Introduction to the Theory of Elasticity}.
\newblock Longman Mathematical Texts. Longman, 1980.

\bibitem{comreddy:1}
B.~D. Reddy.
\newblock By communication.
\newblock {\em University of Cape Town}, 1994.

\bibitem{aruna:1}
B.D. Reddy.
\newblock Stability and convergence of a class of enhanced strain methods.
\newblock {\em SIAM Journal of Numerical Analysis}, 32:1705--1728, 1995.

\end{thebibliography}

\end{document}